\documentclass[preprint]{aastex}

\usepackage{color}
\usepackage{lineno}
\usepackage{rotating}
\usepackage{lscape}

\shorttitle{X-ray variability of GeV gamma-ray emitting radio galaxy NGC 1275}
\shortauthors{Y. Fukazawa et al.}

\begin{document}

\title{X-ray variability of GeV gamma-ray emitting radio galaxy NGC 1275}

\author{Yasushi Fukazawa\altaffilmark{1,2,3}, 
Kensei Shiki\altaffilmark{1}, 
Yasuyuki Tanaka\altaffilmark{2,3}, 
Ryosuke Itoh\altaffilmark{1}, 
and Hiroshi Nagai\altaffilmark{4}}

\email{\texttt{fukazawa@hep01.hepl.hiroshima-u.ac.jp}}

\altaffiltext{1}{Department of Physical Science, Hiroshima University, 1-3-1 Kagamiyama, Higashi-Hiroshima, Hiroshima 739-8526, Japan}
\altaffiltext{2}{Hiroshima Astrophysical Science Center, Hiroshima University, 1-3-1 Kagamiyama, Higashi-Hiroshima, Hiroshima 739-8526, Japan}
\altaffiltext{3}{Core Research for Energetic Universe (Core-U), Hiroshima University, 1-3-1 Kagamiyama, Higashi-Hiroshima, Hiroshima 739-8526, Japan}
\altaffiltext{4}{National Astronomical Observatory of Japan, Osawa 2-21-1, Mitaka, Tokyo 181-8588, Japan}

\begin{abstract}

We analyzed {\em Suzaku}/XIS data of 2006--2015 observations of 
a gamma-ray emitting radio galaxy NGC 1275, and
brightening of the nucleus in the X-ray band was found 
in 2013--2015, correlating with GeV Gamma-ray brightening.
This is the first evidence of variability with correlation between GeV
gamma-ray and X-ray for NGC 1275.
We also analyzed {\em Swift}/XRT data of NGC 1275, and found that
X-ray was flaring by a factor of $\sim$5 in several days in 2006,
2011, and 2013.
The X-ray spectrum during the flare was featureless and somewhat 
steeper with a photon
index of $\sim$2 against $\sim$1.7 in the normal state, indicating that a
synchrotron component became brighter.
A large Xray to GeV gamma-ray flux ratio in the flare could be
 explained by the shock-in-jet scenario.
On the other hand, a long-term gradual brightening of radio, X-ray, and GeV
gamma-ray with a larger gamma-ray amplitude could be origin of other
than internal shocks, and then we discuss some possibilities.

\end{abstract}


\section{Introduction}
NGC~1275 is an elliptical galaxy, locating at the center of the 
Perseus cluster, and is also known to host an AGN and thus 
classified as a radio-loud Seyfert galaxy or a radio galaxy.
Radio galaxies are very important to study jet phenomena.
A viewing angle of their jets is not so small as blazars, and thus the
apparent enhancement of the jet core emission due to the beaming
effect is not so significant and we can probe the jet structure by
seeing the jet periphery emission by observing radio galaxies.

In recent years, {\em Fermi} gamma-ray space telescope detected
GeV gamma-ray emission from radio galaxies \citep{Abdo10}, and 
NGC~1275 is the brightest among them.
Since NGC 1275 was not detected with {\em CGRO}/EGRET and the gamma-ray
flux of {\em Fermi} is much higher than the EGRET sensitivity
limit, gamma-ray emission from NGC 1275 turned out to 
have increased over ten times \citep{Abdo09}. 
{\em Fermi} observation showed the time variation of gamma-ray flux with
several months scale \citep{Kata10} and
the gamma-ray flare was also reported \citep{Dona10, Brow11}.
These results rule out the possibility that the gamma-ray emission comes
from the Perseus cluster via the cosmic ray interactions with the
intracluster medium or the dark matter annihilation.
In addition, 
TeV gamma-rays were also detected from NGC 1275 with MAGIC \citep{Alek12}, 
and a correlated variability between GeV and optical was also 
found \citep{Alek14}.

Therefore, a radio-loud gamma-ray emitting Seyfert galaxy NGC 1275 
is now very attracting to study radio galaxies.
The spectral energy distribution (SED) of NGC 1275 
is reported to be explained by the
synchrotron self-Compton (SSC) model \citep{Abdo09}.
However, the jet emission has not yet been confirmed in the X-ray
band as described below.
Therefore, the SED data of the NGC 1275 jet component currently rely 
mainly on the radio and gamma-ray band, and the SSC model parameters could
change significantly, dependently on the X-ray jet flux.

In the X-ray band, the hosting Perseus cluster is very bright and a
point-like source was for the first time resolved with 
{\em Einstein}/HRI \citep{Bran81}.
{\em XMM-Newton} and {\em Chandra} could also resolve the nucleus emission
spatially \citep{Chur03,Baim06}.
\citet{Fabi15} reported a historical X-ray light curve
of NGC 1275 by collecting the results of many past satellites from
OSO-8 to Swift/XRT.
and showed that X-ray flux was very high in 1980s,
decreased after 1990s, and then are becoming brighter again after 2000s.
Based on the {\em XMM-Newton} observation in 2001 and 2006,
the X-ray spectrum of the nucleus was reported to be represented by the
powerlaw with a photon Index of 1.70--1.75 and 
a flux of $(3-6)\times10^{-12}$ erg cm$^{-2}$ s$^{-1}$ in 5--10 keV 
\citep{Chur03,Yama13}.
However, X-ray and GeV gamma-ray connection and origin of X-rays 
from NGC 1275 nucleus have been not well understood.
\citet{Yama13} reported that the {\em Suzaku}/XIS 
monitoring observations of NGC
1275 in 2006--2011 showed no clear variability against a GeV gamma-ray
variability by a factor of 3.
This indicates that jet emission is not dominated in the X-ray band.

The {\em Fermi} light
curve showed a gradual GeV gamma-ray flux increase after 2010, almost
correlating with 90 GHz radio flux \citep{Duts14}.
VLBI observations confirmed that the radio outburst since 2005 was 
associated with the emergence of a new component \citep{Naga10}, 
and \citep{Naga12} reported that the radio flux increase was mostly attributed
to one compact component C3.
In this paper, we reported the extended analysis of {\em Suzaku}/XIS
data of NGC 1275 up to 2015 after \citet{Yama13},
and also showed systematic analysis of {\em
Swift}/XRT data of NGC 1275 in 2006, 2011, and 2013.
Through this paper, errors correspond to 90\% confidence range.

\section{{\em Suzaku}/XIS Data Analysis and Results}

We performed the X-ray imaging spectroscopy, using the archival 
{\em Suzaku}/XIS \citep{Mits07,Koya07} data of NGC 1275, which was repeatedly observed with
{\em Suzaku} every half year as a calibration target; 19 observations 
from 2006 to 2015 (table 1).
Analysis method is described in detail in \citet{Yama13}, and
hare we extended the {\em Suzaku}/XIS data analysis to 2015.
In the {\em Suzaku}/XIS data, we can see an X-ray enhancement at NGC
1275 against the ambient Perseus cluster emission in the XIS image.
The excess becomes larger in higher energy band up to 12 keV, indicating
that the X-ray spectrum is harder than the cluster emission.
We evaluated this excess emission against the ambient cluster emission
from the radial count rate profile centered on NGC 1275 in 9 energy
bands and derived the X-ray spectra of excess emission in each observation.
The X-ray spectrum of the excess emission is roughly represented by the
absorbed power-law with a photon index of 1.6--1.8, and we obtained
a history of the 5--10 keV flux from the derived spectra.

\begin{figure}[!h]
\begin{center}
\includegraphics[width=12cm]{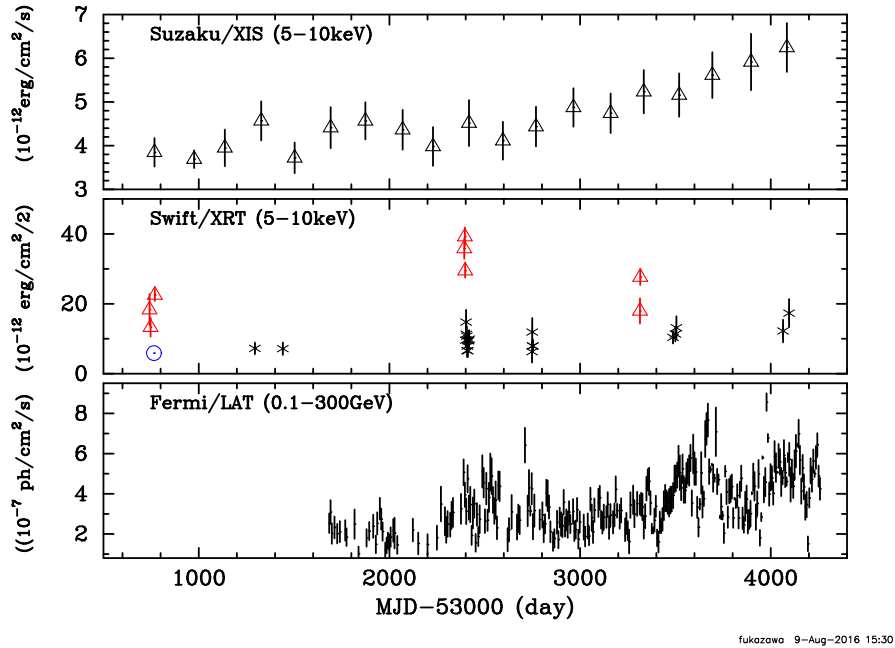}
\vspace*{0.5cm}
\caption{Light curve of the NGC 1275 nucleus. 
(Top) {\em Suzaku}/XIS X-ray  light curve in 5--10  keV,
obtained by the analysis of radial count profiles. 
(Middle) {\em Swift}/XRT X-ray  light curve in 5--10  keV,
obtained by spectral fitting.
Circle is a {\em XMM-Newton} result in 2006.
(Bottom) {\em Fermi}/LAT GeV
gamma-ray light curve in 0.1--300 GeV. 1 bin = 1 week. 
The data are taken from the {\em  Fermi}-LAT archival light curve
({\footnotesize{\tt fermi.gsfc.nasa.gov/ssc/data/access/lat/msl\_lc/}}).}
\label{lc}
\end{center}
\end{figure}

Figure \ref{lc} top shows a {\em Suzaku}/XIS X-ray light curve of NGC
1275 in 2006--2015.
It can be seen that the X-ray flux was constant before 2012 and 
increased gradually since 2013.
X-ray flux increase is a factor of 1.5 between 2012 and 2015.
Figure \ref{lc} bottom plots the archival GeV gamma-ray light curve 
with one week time bin.
We can see a similar behavior between X-ray and GeV gamma-ray light
curves; both light curves show a gradual flux increase since 2013.
This is the first evidence of positive correlation between X-ray and
GeV gamma-ray.

\begin{table}[t]
\begin{center}
\caption{Summary of {\em Suzaku} observations of NGC 1275 in 2012--2015}
%\hspace{-1.8cm}
\begin{tabular}{ccccc}
\hline
\hline
Observation Start & Sequence No. & R.A. Dec. $^{\dagger}$ &  Euler angle & Exposure (XIS)$^{\ast}$ \\ 
& & (deg) & (deg) & (sec) \\
\hline
2012-02-07 20:00:00 & 106005020	& 49.9506 41.4000 & 49.9531 48.4984 188.0014 & 93635  (0,3) \\
2012-08-20 23:30:00 & 107005010 & 49.9347 41.5380 & 49.9448 48.4826 17.3897 & 82263  (0,3) \\
2013-02-11 04:59:00 & 107005020	& 49.9644 41.4892 & 49.9561 48.4897 193.7124 & 82795 (0,3) \\
2013-08-15 10:14:00 & 108005010	& 49.9376 41.5389 & 49.9459 48.4822 13.7581 & 82514  (0,3) \\
2014-02-05 12:30:00 & 108005020	& 49.9630 41.4842 & 49.9546 48.4948 194.0004 & 75999  (0,3) \\
2014-08-27 15:50:00 & 109005010 & 49.9309 41.5377 & 49.9434 48.4822 22.6569 & 40051  (0,3) \\
2015-03-03 17:25:00 & 109005020	& 49.9680 41.4854 & 49.9560 48.4945 201.4727 & 74409  (0,3) \\
\hline
\end{tabular}
\label{suzakulog}
\end{center}
$\dagger$: Observation log in 2006--2011 is given in \citet{Yama13}.\\
$\dagger$: Pointing direction.\\
$\ast$: Exposure time summed over XIS-0, 2, 3. The values in the parenthesis represent the XIS detector ID used for the analysis.
\end{table}

\section{{\em Swift}/XRT Data Analysis and Results}

We analyzed archival {\em Swift}/XRT \citep{Gehr04,Burr05} 
data of NGC 1275 as summarized in table \ref{tab:swift}.
NGC 1275 was observed several times in 2006, 2010, and 2013.
The 2010 and 2013 observations were triggered as ToO, and the 2013
observations were triggered by {\em Fermi} GeV gamma-ray flux increase
\citep{Cipr13}.
Observations were performed in both of the Windowed-Timing (WT) and 
Photon-Counting (PC) modes, as summarized in table \ref{tab:swift}.
We analyzed the data with HEADAS 6.11.
Following the analysis of {\em XMM-Newton} data in \citet{Yama13},
the source spectra was extracted in 0.1 arcmin of NGC 1275.
Note that the PSF of {\em Swift}/XRT telescope is almost the same as that of
{\em XMM-Newton} telescope.
Background spectra are extracted from 60--61 arcsec of the center.
We fitted all the XRT spectra with two-temperature {\tt apec} plus
powerlaw model; {\tt phabs*(apec+apec+powerlaw)} in the {\tt XSPEC} model.
Parameters of the {APEC} model, which represents a thermal emission
from the Perseus cluster, are fixed to the {\em XMM-Newton} values used in
\citet{Yama13}, a temperature of 0.8 keV and a metal abundance
of 0.5 solar.

\begin{figure}[!t]
\begin{tabular}{cc}
\begin{minipage}{0.5\hsize}
\begin{center}
\includegraphics[width=8cm]{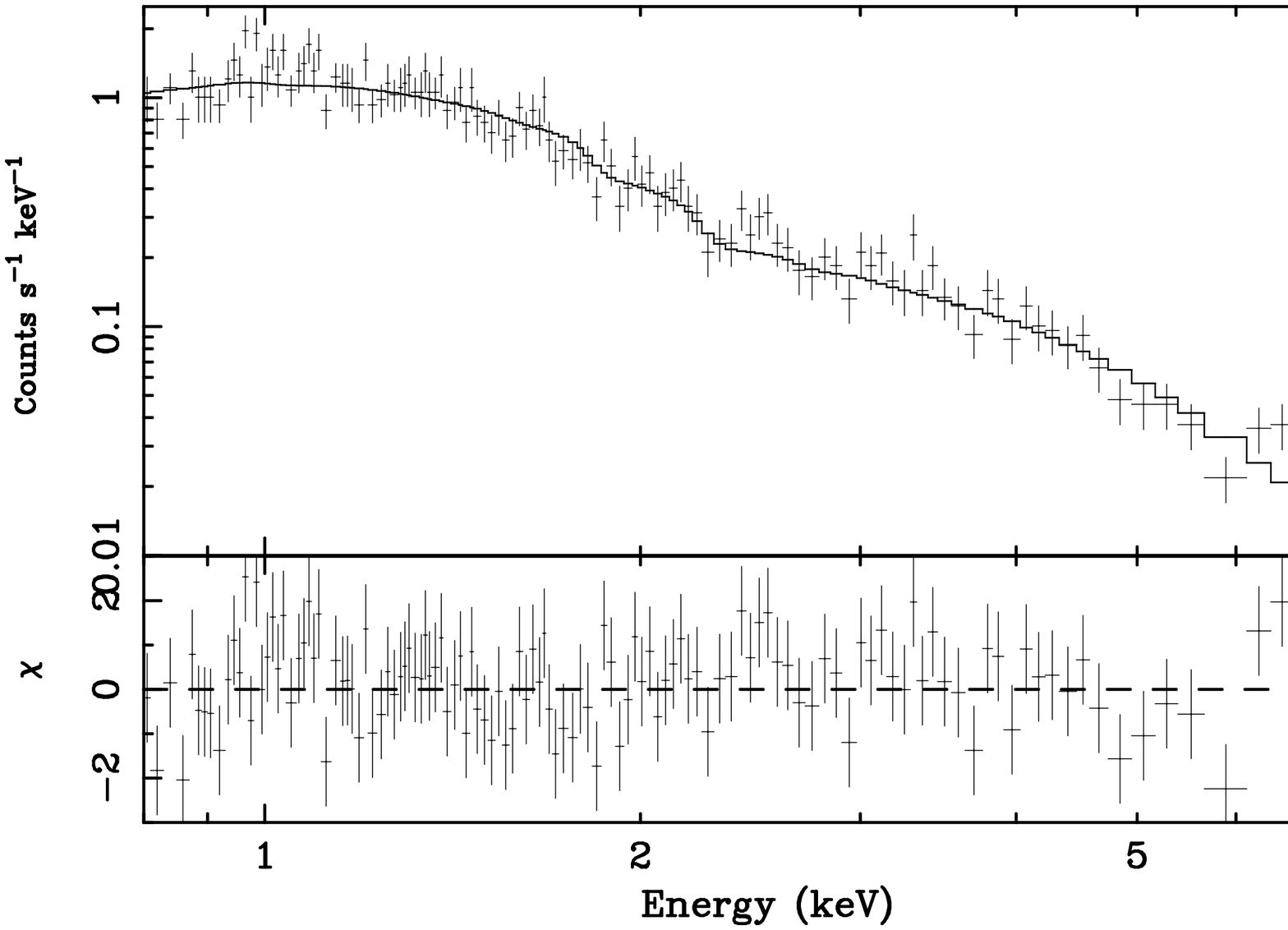}
\end{center}
\end{minipage}
\begin{minipage}{0.5\hsize}
\begin{center}
\includegraphics[width=8cm]{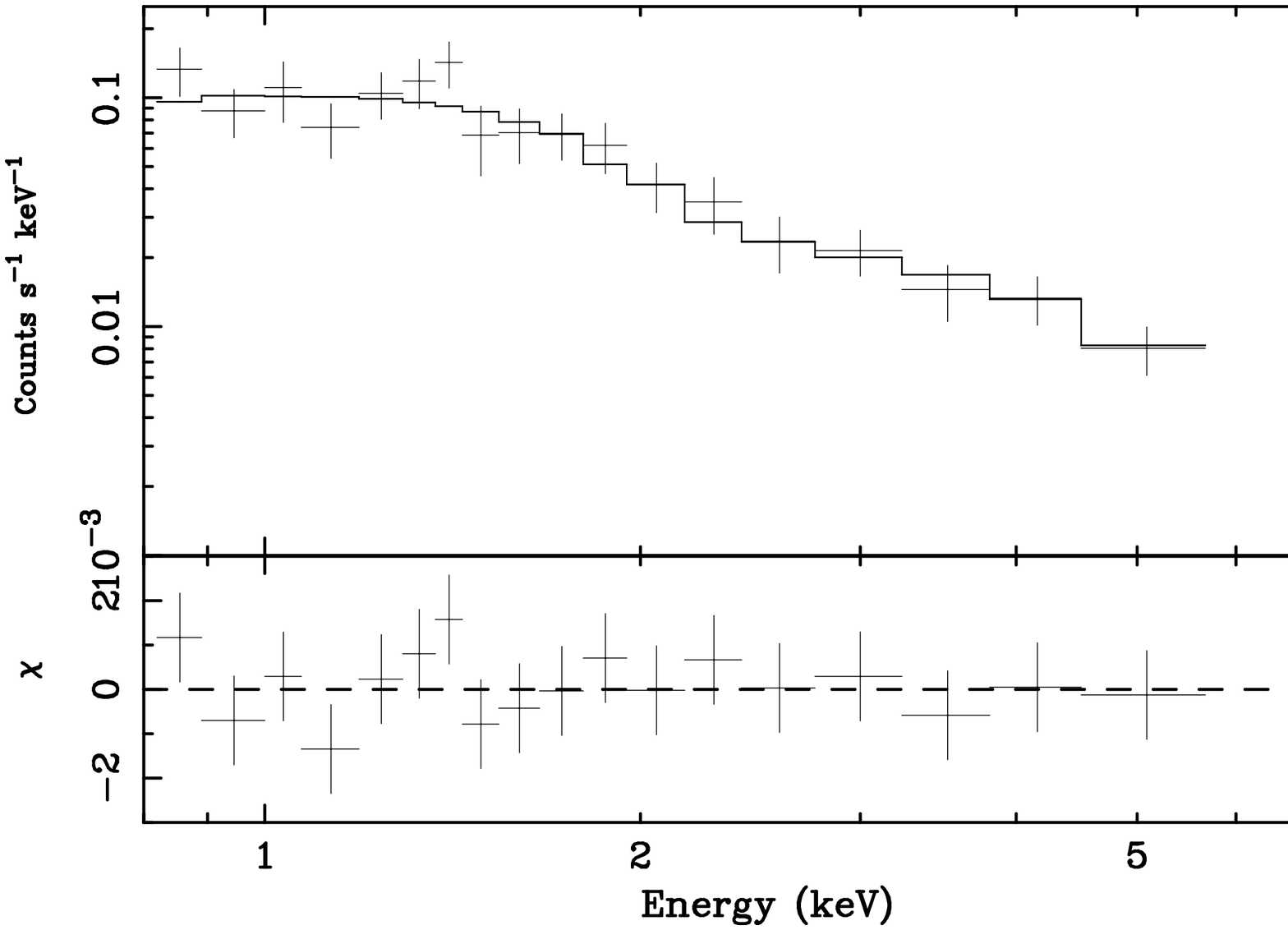}
\vspace*{0.5cm}
\end{center}
\end{minipage}
\end{tabular}
\vspace*{0.3cm}
\caption{{\em Swift}/XRT spectra of NGC 1275 within 0.1 arcmin on MJD=55392
(WT mode, left) and MJD=55403 (PC mode, right). Solid
lines represent the best-fit {\tt wabs*(apec+powerlaw)} model. Bottom
panels show the fitting residuals.}
\label{xrtspec}
\end{figure}

\begin{figure}[!t]
\begin{center}
\includegraphics[width=8cm]{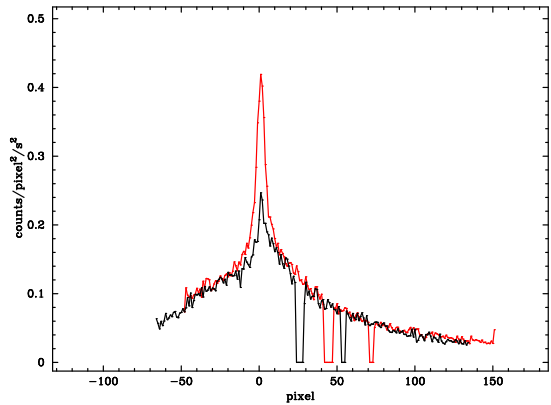}
\vspace*{0.3cm}
\caption{Comparison of count rate profile around NGC 1275. Red and
black lines correspond to the WT (MJD=55395) and PC (MJD=55415) modes, respectively.}
\label{compxy}
\end{center}
\end{figure}

Figure \ref{xrtspec} shows an example of XRT spectra in PC and WT modes.
X-ray spectra in both modes  are featureless, but note that the flux
level is much higher in the WT mode than that in the PC mode.
Figure \ref{compxy} shows the count rate profiles around NGC 1275 for the slit
region of $8\times$8 arcmin$^2$ in the PC mode for the
corresponding region in the WT mode in which onboard processing
projects the data with 8 arcmin width into one line region.
Extended X-ray emission of the ambient intracluster medium is 
the same level within 10\% 
while the nuclear emission is brighter by a factor of $>2$ in the WT mode
than that in the PC mode.

\begin{figure}
\begin{minipage}{0.5\hsize}
\begin{center}
\hspace*{-1.5cm}
\includegraphics[width=8cm]{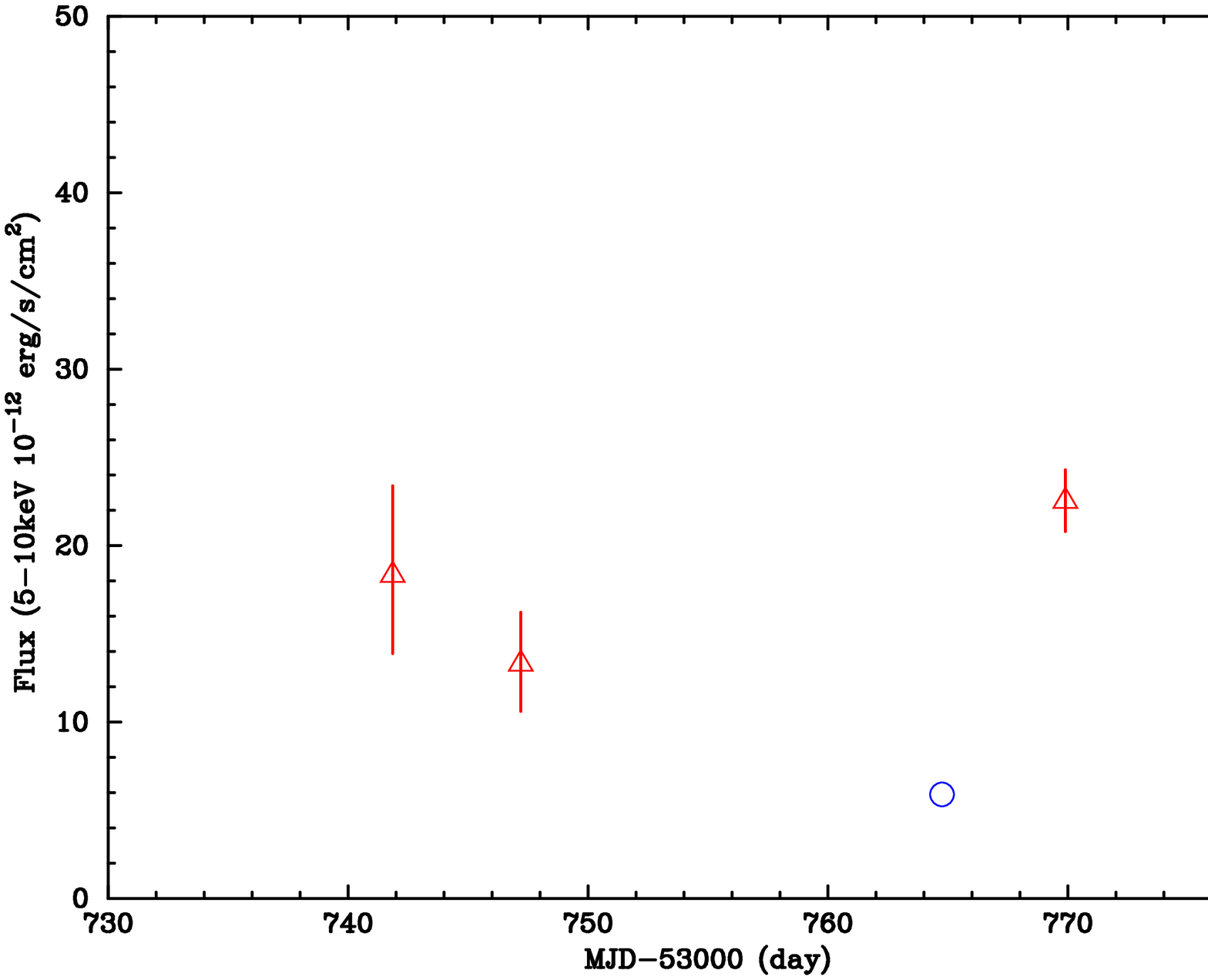}
\end{center}
\end{minipage}\quad
\begin{minipage}{0.5\hsize}
\begin{center}
\includegraphics[width=8cm]{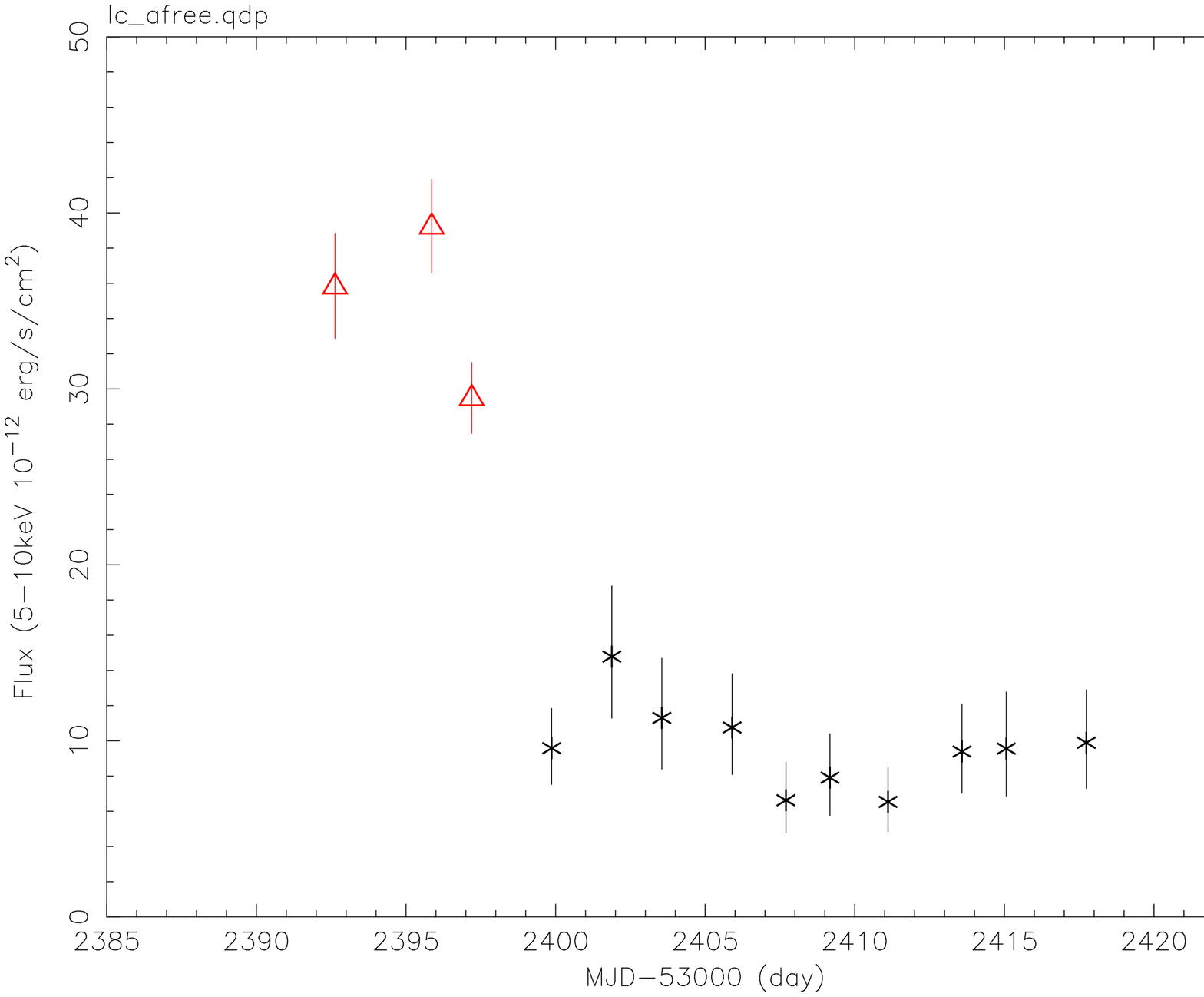}
\end{center}
\end{minipage}
\vspace*{1cm}
\caption{Swift/XRT X-ray light curve of NGC 1275 around MJD=750 (2006) and
MJD=55400 (2010). Circle in the left panel is a {\em XMM-Newton} result in 2006.}
\label{swiftlc}
\vspace*{0.5cm}
\end{figure}

\begin{figure}[!t]
\begin{center}
\includegraphics[width=8cm]{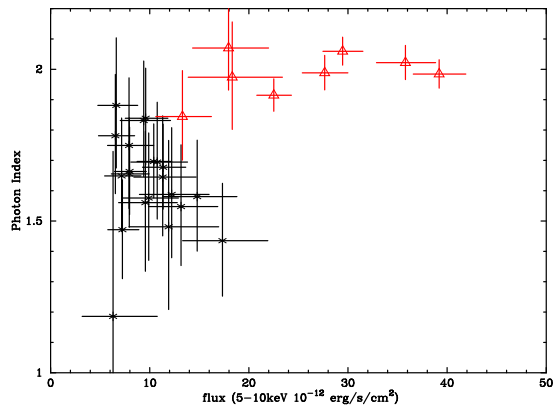}
\vspace*{0.3cm}
\caption{Correlation between the Photon index and the flux (5--10
keV). Circles and triangles are Photon counting and Windows mode, 
respectively.}
\label{a_vs_norm}
\end{center}
\end{figure}

Figure \ref{lc} middle shows a time history of X-ray flux in 5--10 keV.
{\em XMM-Newton} flux in 2006 was also plotted.
The flux observed in the WT mode is very high by a factor of 2--4 
against that observed in the PC mode or observed with {\em XMM-Newton}, 
and hereafter we call them "X-ray flare".
The timescale of X-ray flux change is shorter than 5 days in both 
2006 and 2010.
As shown in figure \ref{swiftlc}, 
the X-ray flare continued in 5--10 days in 2011.
Figure \ref{a_vs_norm} shows a correlation between an X-ray flux 
in 5--10 keV and a photon index.
Photon index during the X-ray flare is around 2, while it is
around 1.5--1.7 during the persistent state, indicating that the X-ray
spectrum became steeper in the X-ray flare.
For the WT data 00031763001, we tried to fit the spectrum by
double-powerlaw model, where one powerlaw has a photon index of 1.7 and
a flux of $4\times10^{-12}$ erg cm$^{-2}$ s$^{-1}$ in 5--10 keV 
and the other one has free parameters.
As a result, the other powerlaw model became a dominant component in the
spectrum and gave a photon index of 2.05$\pm0.06$.
Thefefore, the flaring component has a photon index of 2.0--2.1.

All of {\em Swift}/XRT spectra are featureless, while the XMM-Newton spectra
show a fluorescence Fe-K line.
The tightest upper limit of Fe-K line equivalent width (EW) 
is 619 eV and 191 eV in PC and WT modes, 
and both constraints are consistent with EWs measured with {\em
XMM-Newton}; 
a signal-to-noise ratio of Swift/XRT data is too low to detect a Fe-K
line due to a short exposure.

\begin{table}[t]
\begin{center}
\caption{Summary of {\em Swift} observations of NGC 1275}
%\hspace{-1.8cm}
\begin{tabular}{cccccc}
\hline
\hline
Sequence No. & Observation$^a$ & Exposure$^b$ & Mode$^c$& Flux$^d$ & $\Gamma^e$ \\
\hline
00030354001 & 2006-01-06 (53741) & 1677 & WT & 18.3$_{-4.4}^{ 5.1}$ & 1.97$_{-0.17}^{0.18}$ \\ 
00030354002 & 2006-01-12 (53747) & 3121 & WT & 13.3$_{-2.7}^{ 2.9}$ & 1.84$_{-0.14}^{0.15}$ \\ 
00030364001 & 2006-02-03 (53769) & 3988 & WT & 22.5$_{-1.7}^{ 1.8}$ & 1.91$_{-0.05}^{0.05}$ \\ 
00036524001 & 2007-07-13 (54294) & 5387 & PC &  7.2$_{-1.5}^{ 1.7}$ & 1.47$_{-0.16}^{0.16}$ \\ 
00036524002 & 2007-12-06 (54440) & 3554 & PC &  7.2$_{-1.7}^{ 1.9}$ & 1.65$_{-0.18}^{0.19}$ \\ 
00030354003 & 2009-12-30 (55195) & 4311 & PC &  0.0$_{ 0.0}^{ 0.0}$ & 0.00$_{0.00}^{0.00}$ \\ 
00031763001 & 2010-07-15 (55392) & 1988 & WT & 35.8$_{-2.9}^{ 3.0}$ & 2.02$_{-0.06}^{0.06}$ \\ 
00031763002 & 2010-07-18 (55395) & 4063 & WT & 39.2$_{-2.6}^{ 2.7}$ & 1.98$_{-0.05}^{0.05}$ \\ 
00031763003 & 2010-07-20 (55397) & 4110 & WT & 29.5$_{-2.0}^{ 2.0}$ & 2.06$_{-0.05}^{0.05}$ \\ 
00031770001 & 2010-07-22 (55399) & 2186 & PC &  9.6$_{-2.1}^{ 2.2}$ & 1.84$_{-0.15}^{0.17}$ \\ 
00031770002 & 2010-07-24 (55401) & 2033 & PC & 14.8$_{-3.5}^{ 4.0}$ & 1.58$_{-0.18}^{0.19}$ \\ 
00031770003 & 2010-07-26 (55403) & 2031 & PC & 11.3$_{-2.9}^{ 3.4}$ & 1.64$_{-0.19}^{0.20}$ \\ 
00031770004 & 2010-07-28 (55405) & 2162 & PC & 10.8$_{-2.7}^{ 3.1}$ & 1.69$_{-0.19}^{0.20}$ \\ 
00031770005 & 2010-07-30 (55407) & 2104 & PC &  6.6$_{-1.9}^{ 2.2}$ & 1.88$_{-0.21}^{0.22}$ \\ 
00031770006 & 2010-08-01 (55409) & 2116 & PC &  7.9$_{-2.2}^{ 2.5}$ & 1.75$_{-0.21}^{0.22}$ \\ 
00031770007 & 2010-08-03 (55411) & 2409 & PC &  6.5$_{-1.7}^{ 1.9}$ & 1.78$_{-0.19}^{0.20}$ \\ 
00031770008 & 2010-08-05 (55413) & 1989 & PC &  9.4$_{-2.4}^{ 2.7}$ & 1.83$_{-0.18}^{0.20}$ \\ 
00031770009 & 2010-08-07 (55415) & 2091 & PC &  9.6$_{-2.7}^{ 3.2}$ & 1.56$_{-0.23}^{0.23}$ \\ 
00031770010 & 2010-08-09 (55417) & 2103 & PC &  9.9$_{-2.6}^{ 3.0}$ & 1.58$_{-0.21}^{0.21}$ \\ 
00091128001 & 2011-07-05 (55747) & 1008 & PC &  0.0$_{ 0.0}^{ 0.0}$ & 0.00$_{0.00}^{0.00}$ \\ 
00091128002 & 2011-07-06 (55748) & 1343 & PC &  6.3$_{-3.1}^{ 4.5}$ & 1.19$_{-0.56}^{0.54}$ \\ 
00091128003 & 2011-07-07 (55749) & 1748 & PC & 11.9$_{-4.0}^{ 5.1}$ & 1.48$_{-0.27}^{0.28}$ \\ 
00091128004 & 2011-07-09 (55751) & 3460 & PC &  7.9$_{-1.8}^{ 2.0}$ & 1.66$_{-0.18}^{0.19}$ \\ 
00091128005 & 2011-07-10 (55752) & 4922 & PC &  8.0$_{-1.5}^{ 1.7}$ & 1.66$_{-0.14}^{0.15}$ \\ 
00032691001 & 2013-01-21 (56313) & 1156 & WT & 17.9$_{-3.6}^{ 4.0}$ & 2.07$_{-0.14}^{0.15}$ \\ 
00032691002 & 2013-01-23 (56315) & 2679 & WT & 27.6$_{-2.3}^{ 2.3}$ & 1.99$_{-0.06}^{0.06}$ \\ 
00049799004 & 2013-07-14 (56487) & 5262 & PC & 10.4$_{-1.7}^{ 1.8}$ & 1.70$_{-0.12}^{0.12}$ \\ 
00049799005 & 2013-07-26 (56499) & 3054 & PC & 11.3$_{-2.1}^{ 2.3}$ & 1.68$_{-0.14}^{0.14}$ \\ 
00049799006 & 2013-08-01 (56505) & 1566 & PC & 13.2$_{-3.3}^{ 3.7}$ & 1.55$_{-0.19}^{0.20}$ \\ 
00092034001 & 2015-02-11 (57064) & 1990 & PC & 12.2$_{-3.3}^{ 3.8}$ & 1.59$_{-0.21}^{0.22}$ \\ 
00092034002 & 2015-03-15 (57096) & 2143 & PC & 17.3$_{-4.0}^{ 4.6}$ & 1.44$_{-0.18}^{0.19}$ \\ 
\hline
\end{tabular}
\label{tab:swift}
\end{center}
$a$: Observation Date. MJD is shown in the parenthesis.\\
$b$: PC: Photon-Counting mode, WT: Windowed-Timing mode.\\
$c$: Exposure time in unit of sec.\\
$d$: Flux in 5--10 keV in unit of $10^{-12}$ erg cm$^{-2}$ s$^{-1}$.\\
$e$: Photon index.
\end{table}

%\vspace*{-0.5cm}

%\vspace*{-0.5cm}

%\vspace*{-0.5cm}

\begin{table}[!t]
\begin{center}
\caption{Baseline SED model parameters of NGC 1275 nucleus}
\label{sedpara}
%\vspace{0.2cm}
\begin{tabular}{c|cc} \hline
\hline
 & Base-line & Change \\
\hline
$\Gamma$ & 2.3 & $\times2$ \\
$B$ [G] & 0.035 & $\times2$ \\
$t_v$ [Ms] & $13.4$ & $\times3.3$ \\
$p_1$ & 2.1 &  \\
$p_2$ & 3.1 & -0.2 \\
$\gamma_{\rm min}$ & $8\times10^2$ &  \\
$\gamma_{\rm max}$ & $4\times10^5$ & $\times10$ \\
$\gamma_{\rm brk}$ & $9.6\times10^2$ &  \\
\hline
$P_{j,B}$ [$10^{44}$\,erg s$^{-1}$] & $0.24$ & $\times4$ \\
$P_{j,e}$ [$10^{44}$\,erg s$^{-1}$] & $2.0$ & $\times0.5$ \\
\hline
\end{tabular}
\end{center}
The model parameters are as follows: $\delta$ is the Doppler factor, $B$
 is the magnetic field, $t_v$ is the variability timescale, 
$p_1$ and $p_2$ are the low-energy and
 high-energy electron spectral indices, respectively, $\gamma_{\rm min}$,
 $\gamma_{\rm max}$, and $\gamma_{\rm brk}$ are the minimum, maximum, and break
 electron Lorentz factors, respectively, and $P_{j,B}$ and $P_{j,e}$ are
 the jet powers in magnetic field and electrons,
 respectively.
\end{table}

%\vspace*{-0.5cm}
\begin{figure}[!t]
\begin{center}
\includegraphics[width=12cm]{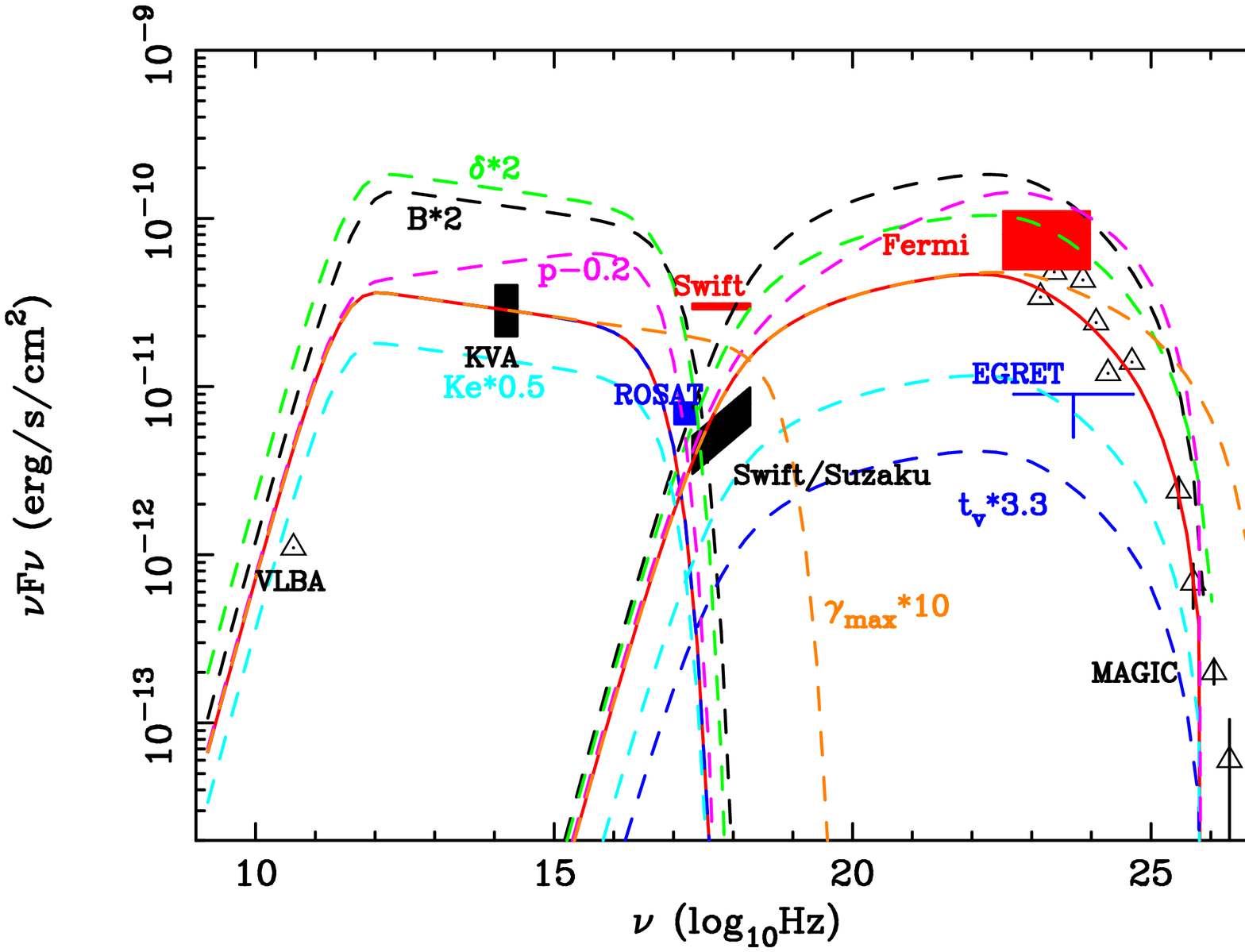}
\vspace*{0.5cm}
\caption{Spectral energy distribution of NGC 1275 variable
 component; 
the VLBA radio data of the C3 component \citep{Suzu12}, optical {\em KVA} data \citep{Alek14}.
{\em ROSAT}/HXI flux \citep{Fabi15}, 
{\em Swift} and {\em Suzaku} in the normal state (this work), 
{\em Swift} flux in the flares (this work).
the gamma-ray data in the normal state
\citep{Abdo09,Alek14}, and the long-term variable GeV gamma-ray flux
(this work).
See the text in detail.
Solid and dashed lines are one-zone synchrotron self-Compton model curves 
calculated with the formula in \citet{Fink08} for various parameter
sets.
Solid red line represents the base line parameter set, and other dashed
lines are curves which are based on the parameter set where only one
 parameter (denoted one) is changed from the base line parameter set. 
See the table \ref{sedpara} in detail.
}
\label{sed}
\end{center}
\end{figure}

\section{Discussion}

Analysis of {\em Suzaku}/XIS and {\em Swift}/XRT data exhibited that
NGC 1275 nuclear X-ray emission varied with a long-term of several
years timescale and a short-term of several days timescale.
During the short-term flare-up, the X-ray spectrum became steeper than
the normal state; the powerlaw photon index became 2.0 from 1.5--1.7
of the normal state.
Figure \ref{sed} shows a SED where the data are
taken from our results and past results.
We plot the data of the long-term variation by black.
For the radio band, we took the VLBA data \citep{Suzu12}.
For the optical data, we took the flux range of {\em KVA} 
light curve in \citet{Alek14}.
For the X-rty data, we plot the {\em ROSAT}/HXI flux \citep{Fabi15},
long-term flux range with a 
photon index of 1.7) of {\em Swift} and {\em Suzaku}
in the normal state (this work), 
and {\em Swift} flux in the flares (this work).
For the gamma-ray, we took the flux in the normal state
\citep{Abdo09,Alek12}, and the high flux state (we scaled the flux in
the normal state by a factor of 2.5, based on figure \ref{lc} bottom).
In the plot, we colored the flare data with red.
Looking at this plot, we discuss the flux variation of NGC 1275.

The long-term X-ray variation was already reported by \citet{Fabi15}
with the past many X-ray satellites, but our results are unique in
terms of showing a positive correlation between X-ray and GeV gamma-ray 
for the first time.
Origin of X-ray emission from the NGC 1275 nucleus is somewhat uncertain.
Several papers assumed that the X-ray emission comes from the inner
jet \citep{Abdo09, Kata10, Suzu12, Alek14}, and obtained the jet
emission parameters by analyzing the SED.
On the other hand, {\em XMM-Newton} spectrum shows a Fe-K line 
with an EW of $\sim$70 eV and a powerlaw
photon index of 1.7 \citep{Chur03, Yama13}.
These properties are typical for Seyfert galaxies, and thus the disk/corona
emission is suggested to be dominant in the normal state.
We suggest that the jet emission largely contributes to the X-ray region
when the GeV gamma-ray flux increases as in 2013--2014.

We plot one-zone synchrotron self-Compton model curves 
calculated with the formula
in \citet{Fink08} for various parameter sets.
We assumed the broken powerlaw for the electron energy distribution as
\citet{Abdo09}.
The base-line parameter set (table \ref{sedpara}) 
is a modification of \citet{Abdo09}; we
adjusted the magnetic field and the variability timescale to be weaker
and shorter, respectively, so that the VLBA, optical, and
X-ray data in the normal state are reproduced.
We calculated the curves by changing only one parameter from the normal
state for magnetic field $B$, Doppler factor $\delta$, variable time
scale $t_{\rm var}$, normalization of electron energy spectrum $K_e$, 
and powerlaw index of
electron energy distribution $p_2$ above the break energy.

In the long-term variation, X-ray flux increase is not as large as GeV
gamma-ray increase.
Such a behavior is reproduced by the increase of electron density $K_e$ or
spectral hardening (decrease of $p_2$) of electron energy 
distribution as shown in figure
\ref{sed} (light blue or purple, respectively).
However, a year-scale gradual hardening of electron energy distribution is
unlikely, 
since the cooling time of high energy electrons is much shorter than one year.
Therefore, the increase of electron density is one possibility.
In addition, the decrease of variable timescale $t_{\rm var}$, in other
word, the reduction of emission region may work together (inferred from the
opposite trend against the blue lines which shows a longer $t_{\rm var}$ case).
We might see the jet collimation for the recent gradual brightening.
On the other hand, 
the radio C3 component brightened since 2003 with an accelerating 
apparent speed of 0.10$c$ to 0.47$c$ \citep{Naga10,Suzu12,Naga12}.
Their flux history is similar to that of GeV gamma-ray and also X-ray.
\citet{Naga12} reported no radio flare during 2010 flare.
Therefore, C3 is likely to be related with the long-term gamma-ray and
X-ray flux increase.
\citet{Naga16} suggested that the C3 is the head of a radio lobe
including a hot spot, and their activity would be enhanced by the
energy input from the nucleus.
As seen in figure 2 of \citet{Naga16}, the radio core (C1 component) has 
also been increasing a little as weel as C3.
Therefore, the increasing activity is one of possibilities of origin of 
long-term gradual increase of  X-ray and gamma-ray flux.

%Using the SSC parameter in table \ref{sedpara}, the energy density of
%electrons becomes $2.4\times10^{-4}$ erg cm$^{-3}$, much higher than the
%pressure of ambient interstellar or intracluster medium.

Here we model the SED by one-zone SSC model, following \citet{Abdo09}.
In this modeling, the X-ray emission in the normal state is low energy
tail of the inverse Compton scattering (IC), and the main IC component
appears above 1 MeV, due to a relatively high minimum
electron energy ($\gamma=$800).
The assumed electron spectrum is as steep as $\propto \gamma^{-3.1}$.
On the other hand, \citet{Alek14} modeled the SSC with a flatter
electron spectrum $\propto \gamma^{-2.55}$ or $\propto \gamma^{-2.4}$ 
and a lower minimum electron energy of $\gamma=100$.
However, their model exceeds the observed X-ray emission in the normal
state.
Therefore, \citet{Tave14} introduced the structured jet model where
the fast spine component main contributes to the low energy peak of SED
while the slow layer surrounding the spine produces the high energy peak of
SED.
At fact, \citet{Naga14} found the limb-brightened structure in the
VLBA core of NGC 1275, suggesting such a structured jet.
In this case, the X-ray emission is a lower part of the main IC component 
coming from the slow layer.

One interesting fact is that NGC 1275 was not detected with {\em
CGRO}/EGRET in 1991--2000, where the X-ray flux is almost the same as
the current flux (ref. X-ray light curve in \citet{Fabi15}).
Current GeV gamma-ray flux observed with {\em Fermi} is brighter than
the EGRET sensitivity.
This indicates that the gamma-ray to X-ray flux ratio was smaller in
1991--2000 by a factor of $>10$ than that in the present days.
The X-ray variation almost follows the radio variation \citep{Fabi15}.
This could be caused by the different electron energy density of
emission region between two epochs; a higher electron energy density in the 
present days by a factor of $>$10 than that in 1990s (light blue).
The difference of emission region size (or $t_{\rm var}$) 
could be origin, but a reduction of emission region size by a factor of
$>3$ from 1990s to the present days is needed, and it seems unlikely.
\citet{Asad06} reported that the radio flux decrease in 1990s is
attributed to the adiabatic expansion of radio lobe in the central
$\sim5$ pc, in agreement with the above scenario.

Short-term flares found with {\em Swift}/XRT coincides with GeV
gamma-ray brightening for the flares in 2010 and 2013 \citep{Brow11,Cipr13}.
This indicates that the origin of X-ray flare is a jet emission.
According to \citet{Brow11}, the 2010 flare showed a GeV gamma-ray flux
increase by a factor of 2--3,
against 4--5 of the X-ray flux increase.
Steeper X-ray spectrum during the flares can be explained by that the
X-ray is a synchrotron high-energy tail unlike the SSC low-energy tail
or part in the normal state.

Another possibility is that the X-ray flare comes from the disk/corona
when the GeV gamma-ray emitting jet emerged.
A radio galaxy 3C 120 shows radio knot ejection events after the dimming
in the X-ray band \citep{Chat09}, and they suggested that the inner 
material of the disk/corona suddenly goes into the central black hole 
and generate a new jet.
At fact, X-ray emission of 3C 120 has been reported to be dominated by
the disk/corona emission \citep{Kata11,Fuka15,Tana15}.
Coincident brightening between X-ray and GeV gamma-ray for NGC 1275
flares cannot be explained by this scenario, and thus the jet origin of
the X-ray flares is likely.

If so, the above features of the X-ray flares, a spectral steeping and a high
X-ray to gamma-ray ratio, can be explained by a higher magnetic
field of the flaring region.
X-ray flares are likely a synchrotron tail, while the synchrotron
emission in the steady state does not reach the X-ray band \citep{Abdo09}.
Stronger magnetic field could move the synchrotron cut-off to higher
energy, and the X-ray flux increases much largely than the gamma-ray
flux does.
However, the X-ray spectrum during the flare is not so steep and thus
the synchrotron cut-off seems to exist in the higher energy band then
the {\em Swift}/XRT band.
Therefore, in addition to the higher magnetic field,
the higher maximum electron energy is needed.
If the maximum electron energy becomes higher by a factor of 10 
in the flare, the synchrotron emission reaches the X-ray band,
as seen in figure \ref{sed} (red dashed line).
Shock-in-jet model can produce such a parameter change.

{\em Hitomi (ASTRO-H)} \citep{Taka14} is very promissing to study the SED of 
NGC 1275 nucleus, 
since it covers a wide X-ray band
from 0.4 keV to 600 keV, and the hard X-ray to soft gamma-ray band is
very sensitive.
Although the sott X-ray band is dominated by the emission from the
ambient intracluster emission,
HXI will resolve the hard core and detect the AGN component above 100 keV.
SGD could detect the core emission if the powerlaw extends beyond 100
keV without break.
If the jet emission dominates above 10 keV, a spectral cut-off feature
typical for Seyfert galaxies is expected to not exist.
In addition, 
SXS could resolve the 6.4 keV line from the 6.7/7.0 keV cluster emission
line, which is very useful for considering the origin of X-ray emission;
we can perform the detail line diagnostics on the line energy, 
line width, Compton shoulder, and so on.
Although the AGN continuum emission would be buried in the intracluster
emission for the SXS spectrum,
due to a large PSF, a simultaneous {\em XMM-Newton} observation will be
very useful to constrain both the 6.4 keV and continuum.
Together with HXI, we can constrain the reflection component of
disk/corona emission and thus obtain the information around the central
engine.

%%%%%%%%%%%%%%%%%%%%%%%%%%%%%%%%%%%%%%%

%%%
% See the manual for the detail.
%%%

\end{document}